\documentclass[aps,prb,showpacs,twocolumn,superscriptaddress]{revtex4-1}

\usepackage{graphicx}
\usepackage{amsmath,amssymb}
\usepackage{hyperref}
\usepackage{natbib}
\usepackage{bm}
\usepackage{color}
\usepackage{mathrsfs}

\begin{document}

\title{Correlated electrons in a Crystalline Topological insulator}

\date{\today}
\author{F.~Manghi}
\affiliation{Dipartimento di Scienze Fisiche, Informatiche e Matematiche, Universit\`a di Modena e Reggio Emilia, Via Campi 213/A, I-41125 Modena, Italy }
\affiliation{CNR - Institute of NanoSciences - S3}

\begin{abstract}

We study the  effects of e-e interaction  in a 3D Crystalline Topological Insulator by adding on-site repulsion to the single-particle Hamiltonian and solving  the many-body problem within Cluster Perturbation Theory. The goal is to clarify how many body effects modify the topological phase that stems from the crystal symmetries. 
Tuning the strength of the on-site interaction we show that band inversion disappears in the bulk and surface states loose their metallic character.

\end{abstract}

\pacs{ 71.10.-w,71.10.Fd ,73.20.-r ,73.43.-f  }

\maketitle
 Crystalline Topological Insulators (CTIs) are a new entry in the fascinating class of topological materials. In these systems, the basic ingredients that give rise to topological insulators - time-reversal invariance and spin-orbit interaction  - are substituted by crystal symmetries. 
After the pioneering work that introduced the very notion of CTIs \cite{Fu},  real materials that exhibit this behaviour have been predicted by theory \cite{Hsieh2012} and revealed by experiments.\cite{Tanaka2012,Dziawa2012,Xu2012,Wang2013,Okada2013,Zeljkovic2014} 
Most of the theoretical studies  for both real  and model CTIs  \cite{Ando2015}  have been based on a single particle description of the electronic states and this in spite of the prediction of  CTI behavior also for materials where strong e-e correlations are expected.\cite{PhysRevB.90.081112,PhysRevLett.112.016403} 
Few works have addressed the effects of electron interactions in CTIs  investigating  how electron correlations affect their topological classifications \cite{Isobe2015} or give rise to new topological crystalline Mott insulator phases.  \cite{Kargarian2013}
The combination of topological band structure and e-e  interactions  has been extensively studied in  \textit{standard} topological insulators (see Refs.~\onlinecite{Assaadrev,doi:10.1142/S0217984914300014,Rachel_2018} for recent reviews).  In this field, the two-dimensional  honeycomb lattice with intrinsic spin-orbit interaction and on-site e-e repulsion - the so-called Kane-Mele-Hubbard model ~\cite{PhysRevB.82.075106}- has been identified as a paradigmatic example. For this system different many-body approaches have been applied, from Quantum Montecarlo simulations ~\cite{PhysRevB.84.205121,PhysRevLett.106.100403,PhysRevB.85.115132} to Quantum Cluster methods,~\cite{PhysRevLett.107.010401,Grandi_2015,Grandi_PRB} showing that the Hubbard
interaction may drastically affect the stability of the quantum spin Hall phase and supress topological protected edge states. 

In this paper we  study the effects of Hubbard correlations on a CTI focusing on quasi-particle states for both bulk and surface systems. We rely our analysis on a solution of the many-body Hamiltonian by Cluster Perturbation Theory(CPT),~\cite{Senechal,Senechal3}  calculating one-particle propagator and spectral functions. 
We investigate the effects of e-e interaction  in the prototypical CTI introduced by Fu \cite{Fu} where a tetragonal unit cell hosts two inequivalent atoms with two orbitals per site.  This simplified structure, containing different sites and different orbitals, may exhibit non-trivial orbital textures that have been shown to be essential in real  CTIs. \cite{Zeljkovic2014,Liu2014,PhysRevB.93.205304,Walkup2018}
We will see that the many-body effects may drastically modify the band orbital order and consequently affect the CTI phase.

This paper is organized as follows: After a  reminder of the single particle properties and an outline of the method adopted to solve the many-body Hamiltonian, the results are shown for both the 3D crystal and the surface-terminated one, illustrating the effects of e-e correlations in terms of orbital order in 3D and of metallic/non-metallic surface states. 

\begin{figure}[h]
\centering \includegraphics[width=8cm]{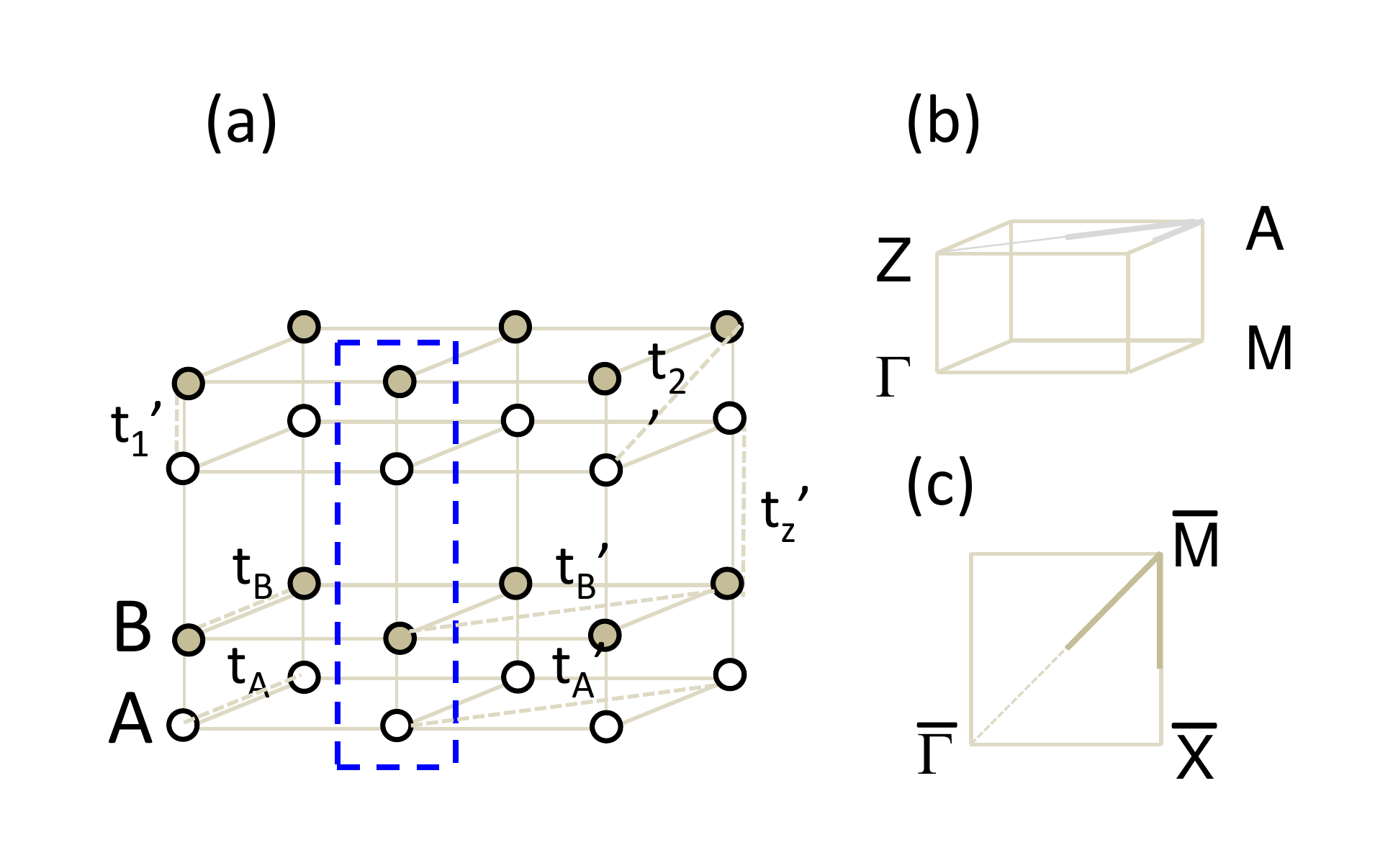}
\caption{\label{geo} (a)  Tetragonal lattice containing two atomic species. Dashed lines enclose  the 4-site non-primitive unit cell  that will be used as the cluster for CPT implementation  (see next section). (b)3D Brillouin  Zone and high symmetry points. (c)2D Brillouin  Zone and high symmetry points. }
\end{figure}

\emph{Single particle band structure}.
As mentioned above, 
the model  consists of a tetragonal lattice  with two atoms, A and B, per cell and two orbitals  per site. Intralayer first- and second- nearest neighbours hopping  parameters connect atoms of the same species with values $  t^A_1=-t^B_1=1, t^A_2=-t^B_2=0.5$; intralayer hopping parameters connect sites A and B  with values $ t_1'=2.5, t'_2=0.5$ , $t'_z=2 $ \cite{Fu}(see (Fig.~\ref{geo} (a)). An orbital dependence is introduced in the intralayer hopping parameters  making them different from zero only when connecting identical orbitals. 

\begin{figure}[h]
\centering \includegraphics[width=9cm]{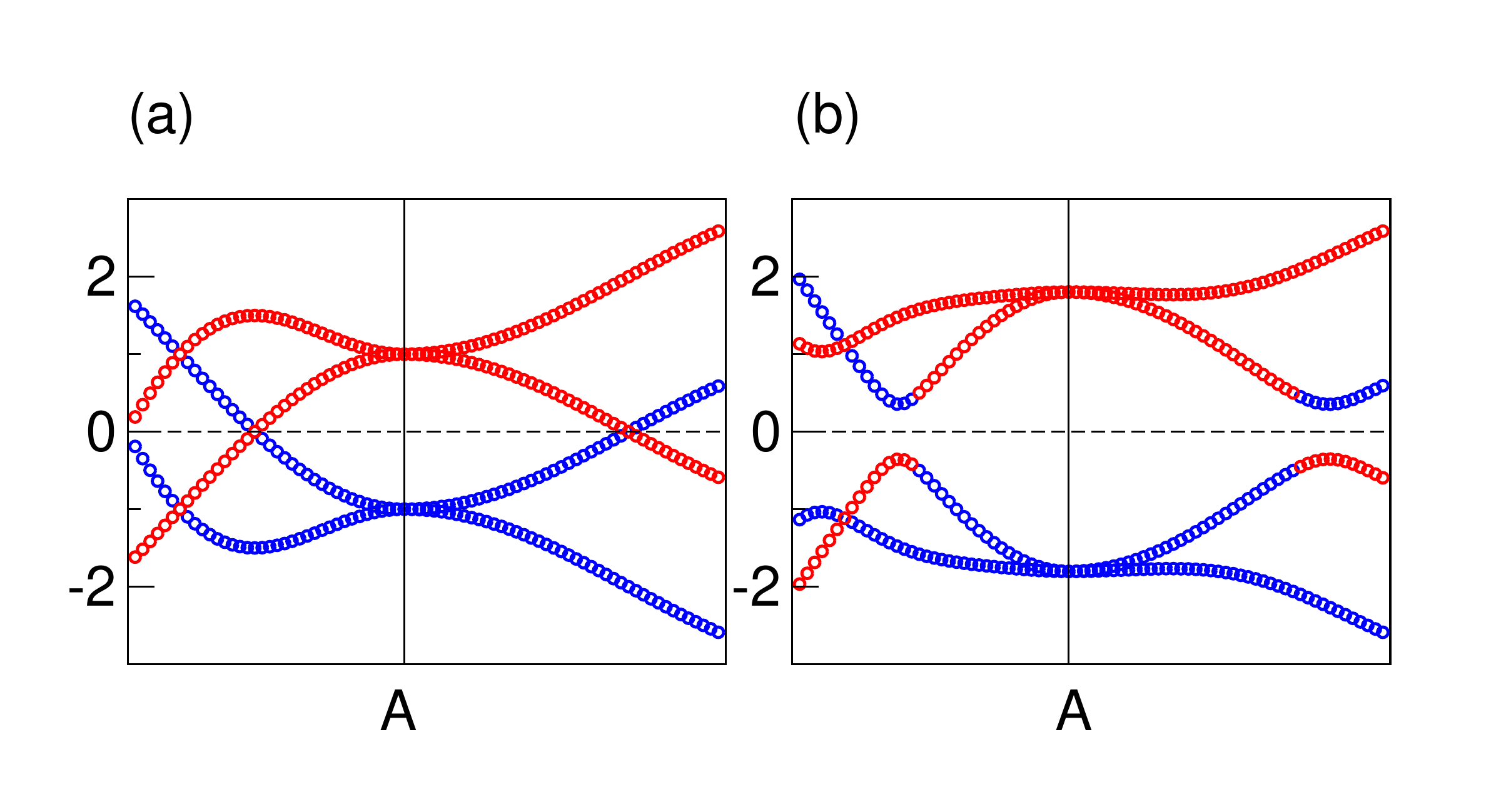}
\caption{\label{ekAB} (a)  Single-particle  band structure calculated at k-points around $A$  where the minimum energy separation between valence and conduction band occurs.   In blue (red) eigenvalues of predominant  A- (B) character. Panel (a) reports eigenvalues obtained with no intralayer hoppings  ($ t_1'=t'_2= t'_z=0 $);  panel (b) eigenvalues obtained with hopping parameters of ref. \onlinecite{Fu}. }
\end{figure}

The 3D crystal is insulating and this thanks to the hopping between sites A and B: indeed, neglecting them, one would get the band structure reported in panel (a) of Fig.~\ref{ekAB} where  bands localized on A sites intersect the corresponding bands associated to B sites. The inclusion of interlayer hopping opens the gap and gives rise to the peculiar orbital texture where states of predominant A and B character interchange along  the valence and conduction band edges (panel (b) of Fig.~\ref{ekAB}). The site composition gives rise to a particular  mechanism of  \emph{band inversion}   essential for  the occurrence of  topologically protected  metallic surface states. Indeed band inversion is ubiquitous in topological materials: it can be associated to orbital  and/or spin degrees of freedom \cite{Wojek2014,PhysRevLett.111.066801,Cao2013,Bernevig1757,PhysRevLett.100.236601,PhysRevLett.111.066801}  or to atomic composition  \cite{Zeljkovic2014,Dziawa2012}. The evolution of band inversion with  e-e interaction will be the main result of this paper.
 
\begin{figure}[h]
\centering \includegraphics[width=9cm]{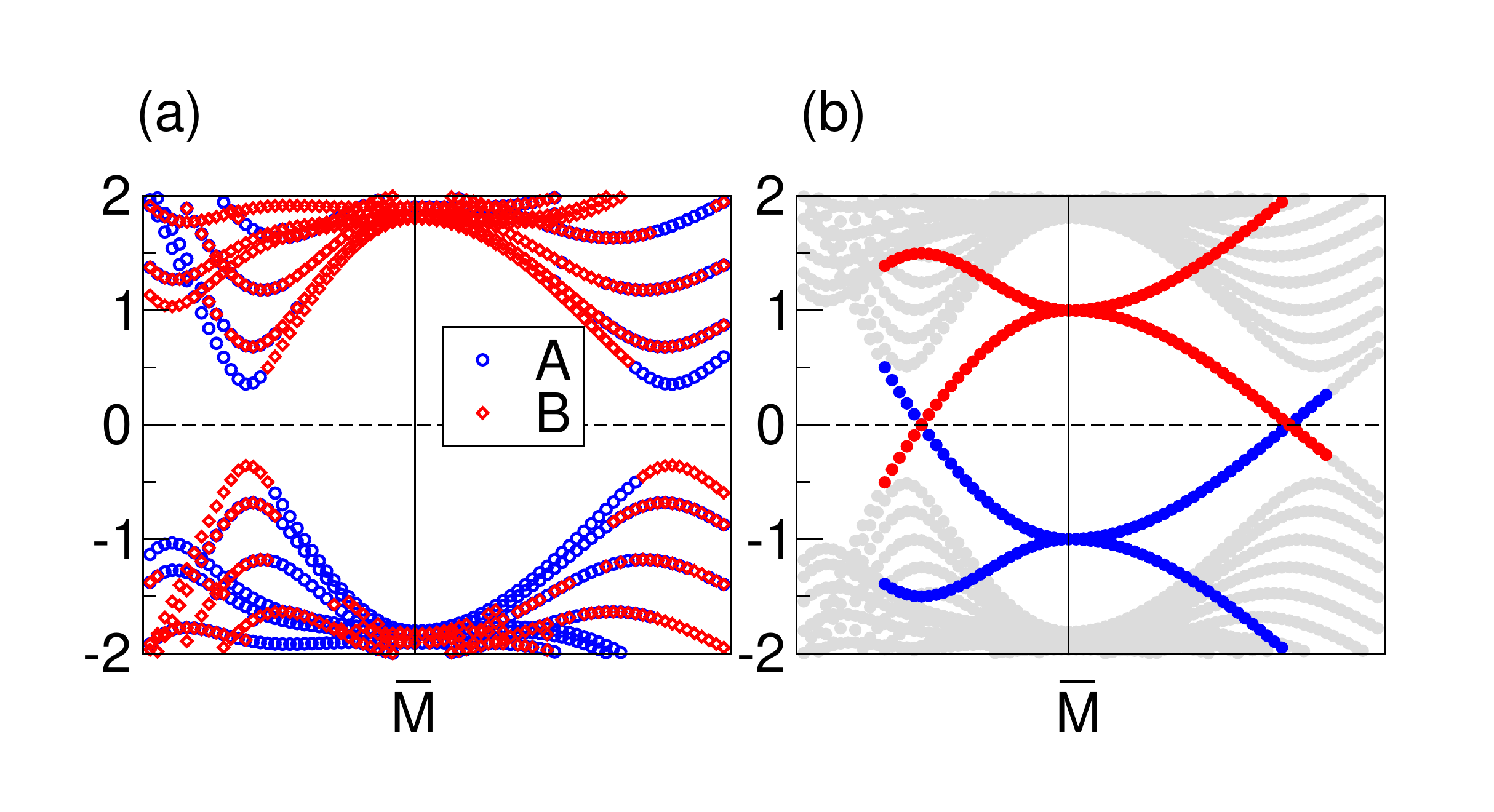}
\caption{\label{SP} a)  Single-particle Projected Bulk Band Structure calculated along high-symmetry directions of the 2D Brilluoine Zone of Fig.~\ref{geo} (c) around $\overline{M}$ point . In blue (red) eigenvalues with predominant A- (B) character. (b) Eigenvalues obtained for a slab of 20  A-B bilayers. In blue (red) eigenstates localized at the lower (upper) surface layer containing A (B) sites. }
\end{figure}

In order to study the reduced dimensionality we have adopted a slab geometry by stacking 20 A-B bilayers along the (001) axis; in this configuration we have both A- and B-terminated surfaces and correspondingly surface states localized either on A or B sites ( Fig.~\ref{SP} (b) ). 
The same slab geometry can be used as a non primitive unit cell to calculate bulk states, getting the so-called Projected
Bulk Band Structure (PBBS) (Fig.~\ref{SP} (a) ). As currently done in  surface physics \cite{Manghi1987} PBBS
allows to identify straightaway the energy
regions that, prohibited in the bulk, can host localized
states at the surface. The orbital texture of PBBS whereby the A and B bands alternate across the gap edges is also shown.
\\

\emph{Interacting Crystalline Topological Insulator}

The interacting Hamiltonian reads: 
\begin{equation}
\label{hubbard}
\hat{H} =  \sum_{\alpha \beta } \sum_{ij l l'}
t_{i l \alpha, j l \beta} \hat{c}_{il \alpha }^{\dagger}
\hat{c}_{jl'  \beta }
 + U \sum_{i l \alpha \beta} 
 \hat{n}_{i l\alpha\uparrow} \hat{n}_{i l \beta \downarrow} 
\end{equation}
Here  $\alpha$,$\beta$ are orbital indices, $i$,$j$ run over the atomic positions within the unit cell  and $l$,$ l'$   refer to lattice vectors. $U$, the strength of the on-site Hubbard interaction, is assumed to be site and orbital independent.

We solve the many-body problem by  Cluster Perturbation Theory  (CPT).~\cite{Senechal}
CPT has been successfully used to study Mott-Hubbard physics both in model systems \cite{Senechal,Potthoff2003} and in real materials,~\cite{ManghiMnO,Eder2008,Eder2015} and more recently to address correlated topological phases of matter.~\cite{PhysRevLett.107.010401,Grandi_2015,Grandi_PRB,PhysRevB.94.161111}

CPT   belongs to the class of Quantum Cluster theories which solve  the problem of many interacting electrons in an extended lattice by  approaching first  the many body problem in a subsystem of finite size - a cluster- and then embedding it within  the infinite medium. \cite{RevModPhysQC}  The lattice is  seen as a periodic repetition of  identical clusters  each of them containing $M$
atoms  characterized by a set of $n_i^{orb}$  orbitals 
(K= $\sum_i^M n_i^{orb}$ is
the total number of
sites/orbitals per cluster).
The  Hamiltonian  is then
partitioned in two terms, an intra-cluster and an inter-cluster one. 
Since the e-e Coulomb interaction is  on-site, the inter-cluster Hamiltonian contains only single particle hopping terms and the  many body part is present in the intra-cluster Hamiltonian only. The Green's function for the extended lattice can then be obtained by solving the equation
\begin{equation}\label{QC}
G_{i j \alpha \beta}(k,   \omega)=G^c_{ i \alpha j \beta}(  \omega)+ \sum_{i' \delta}B_{i \alpha i'\delta}(k,  \omega) G_{i' \delta j \beta}(k,   \omega).
\end{equation}
Here the $K\times K$  matrix $B_{i \alpha i' \delta}(k,\omega)$ is given by
\begin{align*}
B_{i  \alpha i' \delta}(k,\omega)= \sum_l^L e^{i k\cdot R_l} \sum_{i''}^M \sum_{\gamma} G^c_{i \alpha i'' \gamma}(\omega) t_{i'' \gamma 0, i' \beta l}
\end{align*}
with $R_l$ the lattice vectors and $t_{i''0 ,i' l}$  the hopping  between site $i'$ and $i''$ belonging to different clusters.
$G^c_{ i \alpha j \beta}$ is the  Green's function obtained by exact diagonalization of the interacting Hamiltonian for the finite cluster; we separately solve the   problem for  N, N-1 and N+1 electrons and
express the cluster Green's function in the Lehmann representation at real frequencies.
\footnote{More details on CPT implementation for multi-orbital systems can be found in https://www.cond-mat.de/events/correl16/manuscripts/manghi.pdf}

For the tetragonal lattice, the
4-site cluster  shown in Fig.~\ref{geo} (a) has been adopted. This choice is  suggested by the strength of hopping parameters: 
in the present model the largest hopping parameters are those connecting A-B sites. They are included in the exact diagonalization while the smaller  ones are used, in the spirit of CPT,  \emph{perturbatively} in the periodization of the cluster Green's function. 

Once the cluster Green's function in the local basis  $ G^c_{i \alpha i' \beta}(  \omega)$ has been calculated  by exact diagonalization, eq. \ref{QC} is solved   by  matrix inversion at each $k$ and $\omega$. The quasi-particle band structure is then obtained in terms of  spectral function $A(k \omega)$
\begin{equation}
\label{akn}
A(k  \omega) = \frac{1}{\pi}\sum_n Im  G(k n \omega).
\end{equation}
where  
\begin{align*}
G(k n \omega) =  \frac{1}{K}\sum_{i i'} e^{-i k \cdot(r_i-r_{i'})} \mathcal{C}^n_{i \alpha}(k)   G_{i i'}(k,\omega)
 \end{align*}
Here $n$ is the band index and $\mathcal{C}^n_{i \alpha}(\textbf{k})$ are the eigenstate coefficients obtained by the single-particle band calculation.\cite{ManghiMnO}
\\

\emph{Results}.
We compare now the spectral function obtained for the bulk crystal with increasing values of $U$.  We see that the minimum
energy separation between hole and particle excitations - the quasi-particle gap $\Delta$ - diminishes with $U$ and goes to zero at a critical value $U_c=2.5$. After that, $\Delta$ keeps increasing linearly as in a standard  Mott-Hubbard regime. This behaviour resembles quite closely what happens in the Kane-Mele-Hubbard model which describes graphene as a Quantum Spin Hall topological insulator in the presence of on-site e-e interaction \cite{PhysRevLett.107.010401,Grandi_2015,Grandi_PRB,Rachel_2012}.

\begin{figure}[h]
\centering \includegraphics[width=8cm]{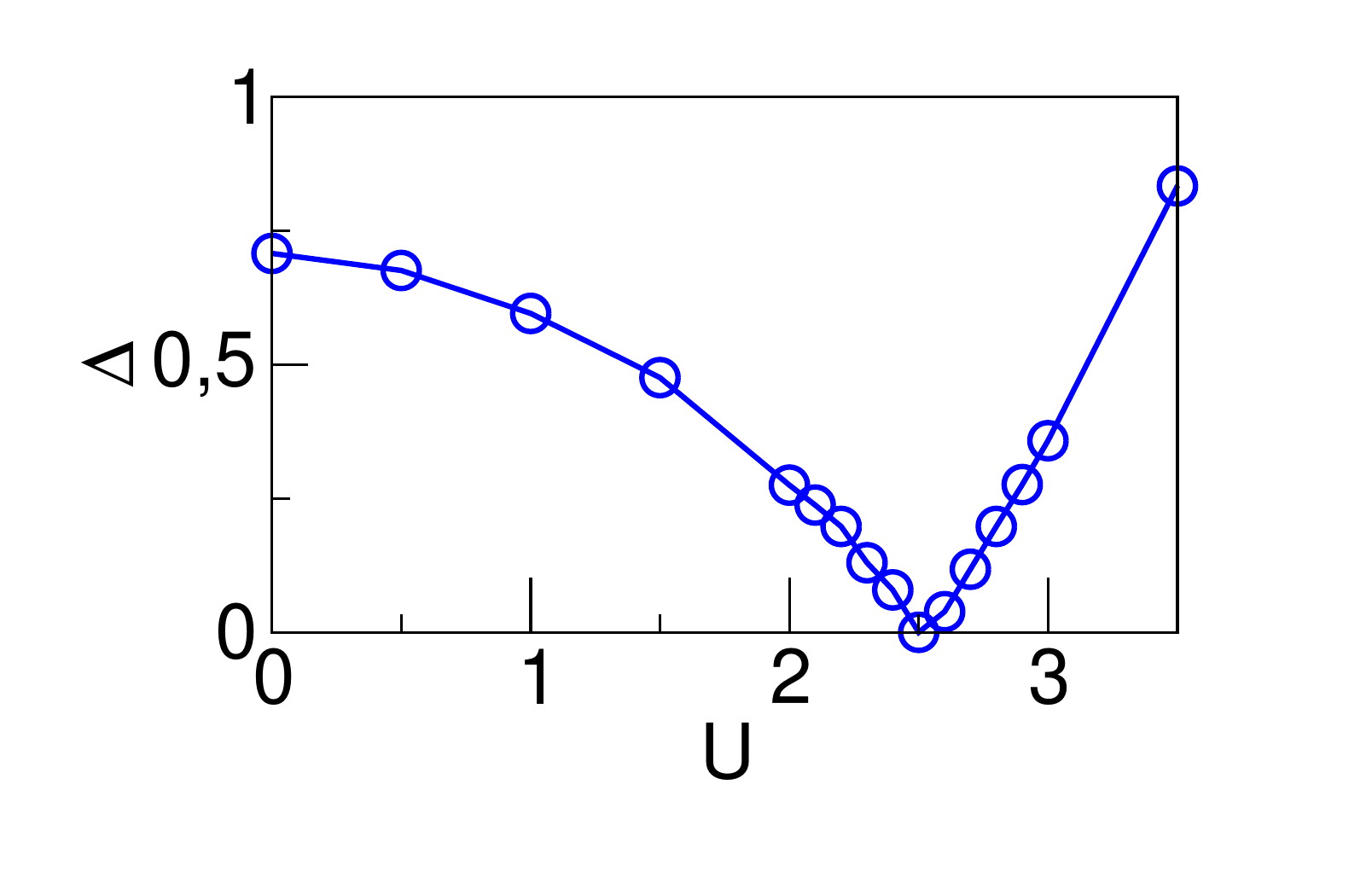}
\caption{\label{gap} Evolution of the minimum energy separation between hole and particle excitations as a function of the strength of the $U$-parameter}
\end{figure}

\begin{figure}[h]
\centering \includegraphics[width=8cm]{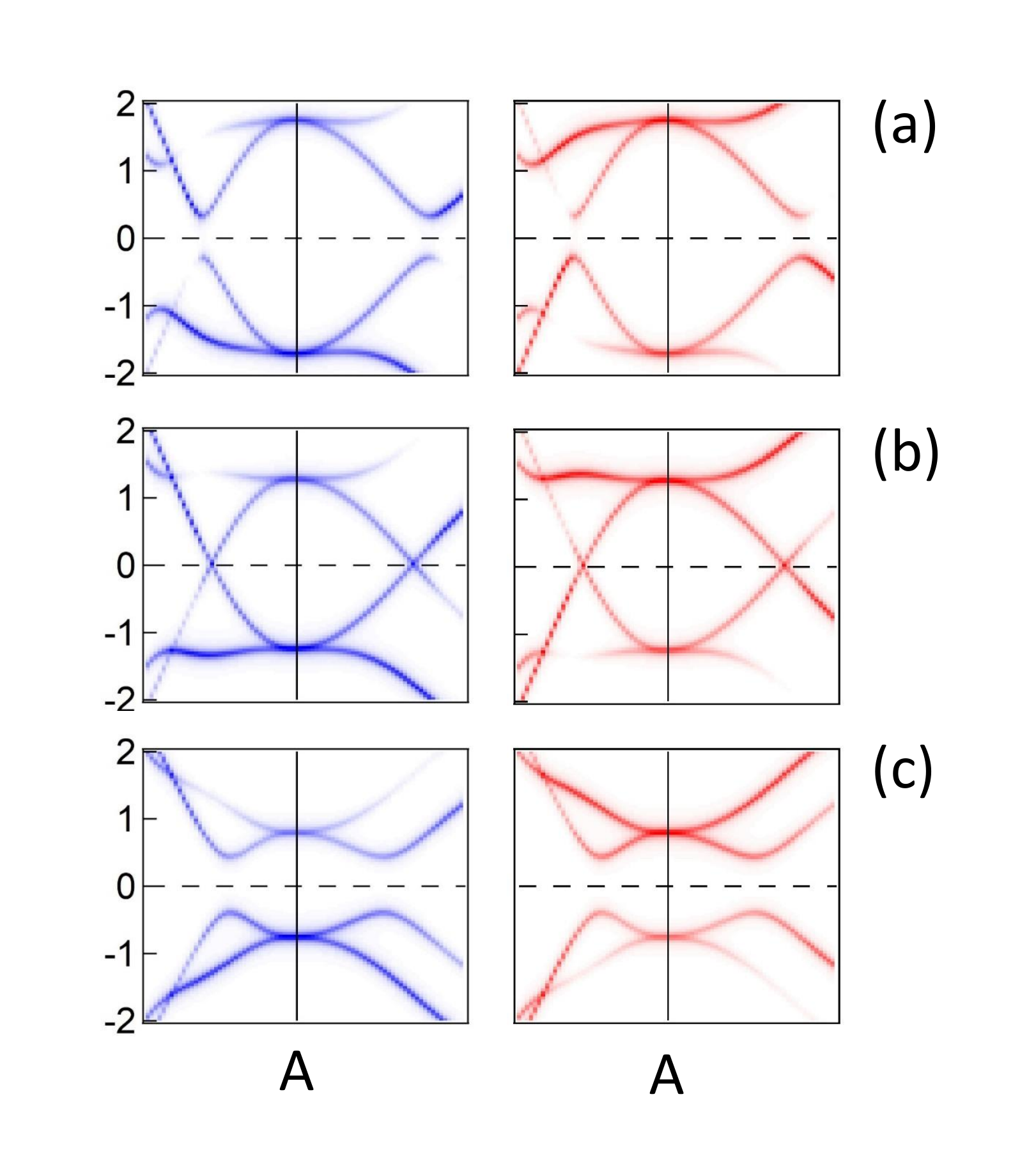}
\caption{ \label{bulkAB} Site selected bulk quasi-particle states at k-points where the minimum energy separation between hole and particle states occurs.  Panels  (a) ,(b), (c) correspond to  $U=1,2.5,3.5 $ respectively. The left (right) panels in blu (red) show the $A$-site ($B$-site) contribution.}
\end{figure}

Fig~\ref{bulkAB} shows the quasiparticle band structure of  bulk CTI  for different values of $U$ around the critical value $U_c$. In order to elucidate the orbital composition of the quasi-particle states we plot 
the \emph{local} spectral function $A(k  \alpha \omega) = \frac{1}{\pi}\sum_n Im  G(k n \omega)|\mathcal{C}^n_{i \alpha}(k)^2|$ with $\alpha$ running over the two orbitals of site $A$ (right panel) or $B$ (left panel). We see that for $U< U_c $ states close to the Fermi level have a predominat $A$-site or $B$-site character. 

For  $U\geq U_c $ this is no more the case and the  states that are close to the Fermi level exhibit a mixed $A$-$B$ composition.  
The orbital texture in the bulk evolves then with $U$: below $U_c$ valence and conduction band edges exhibit the same band inversion that has been identified in the non-interacting system with the predominant contribution from $A$ and $B$ sites alternating around the $A$ point (or the $\overline{M}$ point in the PBBS). This band inversion fades away for $U\geq U_c$ where states close to the Fermi level have a mixed $A$-$B$ character.

\begin{figure}[h]
\centering \includegraphics[width=8cm]{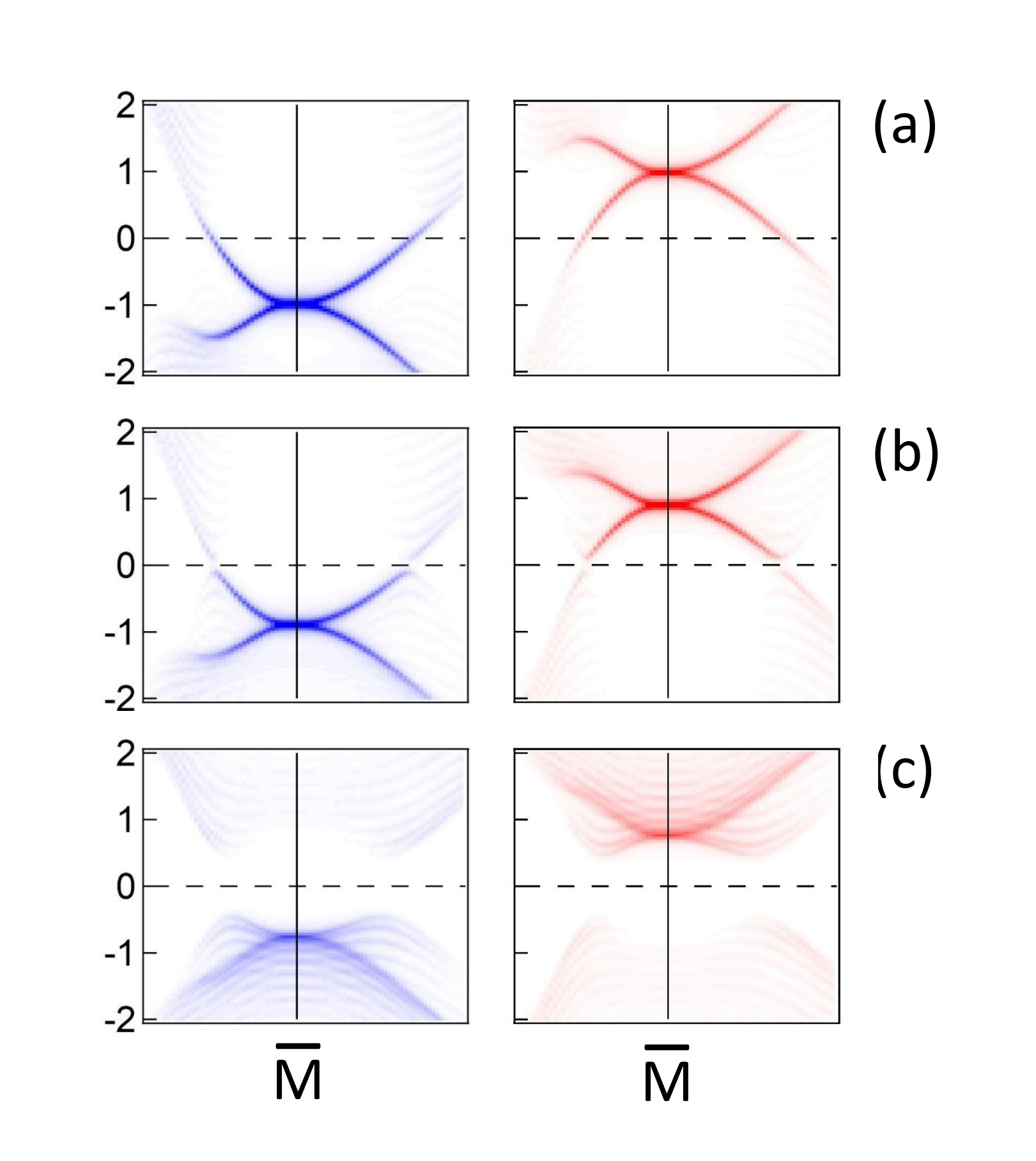}
\caption{ \label{surfAB} Surface quasi-particle states for increasing $U$ across $U_c$ . Panels  (a) ,(b), (c) correspond to  $U=1,2.5,3.5 $ respectively. The left (right) panels in blu (red) show states localized at the $A$-site ($B$-site) terminated surface.}
\end{figure}

Simultaneously, as band inversion disappears, surface states loose their metallic  character  and a gap appears in their k-dispersion. This is shown in Fig.~\ref{surfAB} where the local spectral functions obtained for the 20-layer slab are shown for the same $U$ values. In the non interacting CTI the existence of gapless surface states has been associated to a topological phase; we see that this phase persists only up to $U<U_c$. 

In conclusion, we have demontrated that the Hubbard interaction modifies the orbital texture of the quasi-particle band edges around the Fermi level; this effect is remarkable and can be relevant also in real materials exhibiting CTI behaviour. Above the critical value $U_c$  band inversion is removed and surface states are gapped. This suggests the identification of this critical value as a    transition point from a topological to a trivial insulating phase. 
An explicit confirmation of topological character and of its evolution with $U$ would require the generalization to the interacting case  of the new $Z_2$ topological invariant that has been introduced  to classify the
topological crystalline insulator phase in the non interacting system.~\cite{Fu}  The  so-called \emph{topological Hamiltonian} \cite{PhysRevB.85.165126,PhysRevX.2.031008} - a fictitious non-interacting Hamiltonian associated to the inverse of the dressed Green's function at zero frequency - would be the most promising approach. It has  been applied  to identify the topological character of heavy fermion mixed valence compounds~\cite{PhysRevB.88.035113,PhysRevLett.111.176404,PhysRevLett.110.096401,PhysRevB.87.085134} and of the half-filled honeycomb lattice~\cite{Grandi_2015,Grandi_PRB} in the presence of on-site Hubbard interaction. The application of this method to CTI's will be the subject of a forthcoming paper.

 \end{document}